\documentclass[
    ,final            
  ]
  {aipproc}

\layoutstyle{6x9}
\usepackage{amssymb}

\renewcommand\XFMtitleblock{%
  \XFMtitle
  \let\XFMoldpar\par
  \def\par{\XFMoldpar\def\par{\space 
             (on behalf of the MAGIC Collaboration)\XFMoldpar}}%
   \XFMauthors
   \let\par\XFMoldpar
   \XFMaddresses
   \XFMabstract
   \vspace{5pt}%
   \XFMkeywords
   \XFMclassification
 }

\begin{document}
 
\title{Status and recent results of MAGIC}

\classification{95.85.Pw 95.85.Ry}
\keywords      {Very High Energy Gamma-ray astronomy; MAGIC telescope}

\author{Javier Rico\footnote{Presenter}\ }{
  address={Instituci\'o Catalana de Recerca i Estudis Avancats \& 
Institut de F\'{\i}sica d'Altes Energies;
08193 Bellaterra (Barcelona) Spain }
}
\author{Robert Wagner}{
  address={Max-Planck-Institut f\"ur Physik; D-80805 M\"unchen, Germany}
}

\begin{abstract}

MAGIC is a single-dish Cherenkov telescope located on La Palma
(Spain), hence with an optimal view on the Northern sky. Sensitive in
the 30 GeV -- 30 TeV energy band, it is nowadays the only ground-based
instrument being able to measure high-energy $\gamma$-rays below 100
GeV. We review the most recent experimental results obtained using
MAGIC.
\end{abstract}

\maketitle


\section{Introduction: the MAGIC telescope} 

The Major Atmospheric Gamma Imaging Cherenkov (MAGIC) telescope is a
last-generation instrument for very high energy (VHE) $\gamma$-ray
observation exploiting the Imaging Air Cherenkov (IAC) technique. This
kind of instrument images the Cherenkov light produced in the particle
cascade initiated by a $\gamma$-ray in the atmosphere. Located on the
Roque de los Muchachos Observatory, in La Palma (Spain), MAGIC
incorporates a number of technological improvements in its design and
achieves the lowest energy threshold (55~GeV with the nominal trigger,
25~GeV with the pulsar trigger~\cite{crabpulsar}) among instruments of
its kind. MAGIC signal digitization utilizes 2\,GSample/s Flash
Analog-to-Digital Converters, and timing parameters are used during
the data analysis~\cite{timing}, yielding a sensitivity (at a flux
peak energy of 280\,GeV) of 1.6$\%$ of the Crab Nebula flux in 50
hours of observations.The relative energy resolution above 200~GeV is
better than 30$\%$. The angular resolution is $\sim 0.1^\circ$, while
source localization in the sky is provided with a precision of $\sim
2'$. MAGIC is also unique among IAC telescopes by its capability to
operate under moderate illumination~\cite{moon} (i.e.\ moon or
twilight). This allows to increase the duty cycle by a factor 1.5 and
a better sampling of variable sources is possible. The construction of
a second telescope is now in its final stage and MAGIC will start
stereoscopic observations in the following weeks.

MAGIC has been operating since fall 2004, developing a physics program
which includes both, topics of fundamental physics and
astrophysics. In this paper we highlight MAGIC latest contributions to
Extragalactic
\cite{MAGICM87,alb07a,alb07b,tag,mkn180,MAGIC1011,systHBL,atel1500,MAGICScience}
and Galactic
\cite{crabpulsar,tev,casA,ic443,wr147,cygX1,lsi,crab,psr1951,sgrA,hess1834,hess1813,lsiperiodic,lsimw}
astrophysics.

\section{Extragalactic Observations} 
Except for the radio galaxy M\,87, all 23 currently known VHE
$\gamma$-ray emitting AGNs \cite{Wagner3} are high-frequency peaked
BL~Lac objects\footnote{See {\tt
http://www.mpp.mpg.de/$\sim$rwagner/sources/} for an up-to-date source
list.}. Their two-bumped spectral energy distributions (SED) are
characterized by a second peak at very high $\gamma$-ray energies. In
synchrotron-self-Compton (SSC) models this peak is assumed to be due
to the inverse-Compton (IC) of electrons, by upscattering previously
produced synchrotron photons to high energies. In hadronic models,
instead, interactions of a highly relativistic jet outflow with
ambient matter, proton-induced cascades, or synchrotron radiation off
protons, are the origin of the high-energy photons. Another defining
property of blazars is the high variability of their emission ranging
from radio to $\gamma$-rays. For VHE $\gamma$-ray blazars,
correlations between X-ray and $\gamma$-ray emission have been found
on time scales ranging from $\sim 10$ minutes to days and months (see,
e.g., \citet{fossati}), although the correlations have proven to be
rather complicated \cite{Wagner4}.

Here we present selected results from multi-wavelength (MWL) campaigns
with MAGIC participation and for observations of the AGN
Mkn~501, M\,87 (February 2008), 1ES\,1011+496, S5\,0716+71, and
3C\,279.

\subsection{Multi-Wavelength Campaigns}
For an advanced understanding of blazars, coordinated simultaneous
MWL observations are essential, as they allow the determination 
of SEDs spanning over 15 orders of magnitude in energy. In the 2006/7 
season, MAGIC participated in multiwavelength-campaigns carried out on 
the blazars Mkn 421, Mkn 501, PG 1553+113, 1ES 1218+304, 1H 1426+428, 
and the radio galaxy M\,87. These campaigns involved the 
X-ray satellites {\it INTEGRAL}, {\it Suzaku} and
{\it Swift}, the $\gamma$-ray telescope H.E.S.S., VERITAS, and MAGIC, and
other optical and radio telescopes. We discuss some of the campaigns in 
more detail here. 

The observations of Mkn\,501 in July 2006 revealed the lowest X-ray
and VHE state ever observed. No variability in VHE $\gamma$-rays was
found, while an overall increase of about 50\% during one day was seen
in X-rays. A one-zone SSC model describes this quiescent state of
Mkn\,501 well \cite{mahaya}. Campaigns on 1ES\,1218+304 and 
1H\,1426+428 have been carried out, during both of which significant 
X-ray variability has been observed. Mkn 421 was observed together 
with {\it INTEGRAL}, {\it Suzaku} and {\it Swift} and showed interesting 
intra-day variability during some of the observation nights. The VHE data 
of these observations are currently being analyzed. 
A more recent observation of a flare of Mkn 421 in summer 2008 could be 
followed accurately from optical to VHE $\gamma$  rays, with the 
participation of AGILE, GASP-WEBT, VERITAS and MAGIC \cite{donnarumma}. During 
the first multi-wavelength campaign ever on PG 1553+113 in July 2006, its 
VHE emission showed no variability \cite{reimer}. 1ES\,1959+650 
showed VHE data among the lowest flux state observed from this object, while 
at the same time a relatively high optical and X-ray flux (both {\it Suzaku} 
and {\it Swift}) was found \cite{tag}. The SED could be modeled using similar
parameters as needed for the SED measured in 2002, with a slightly more compact
source and a slightly lower magnetic field.

\subsection{Strong Flaring of Messier\,87 in February 2008}
\begin{figure}
 \resizebox{.85\textwidth}{!}
  {\includegraphics{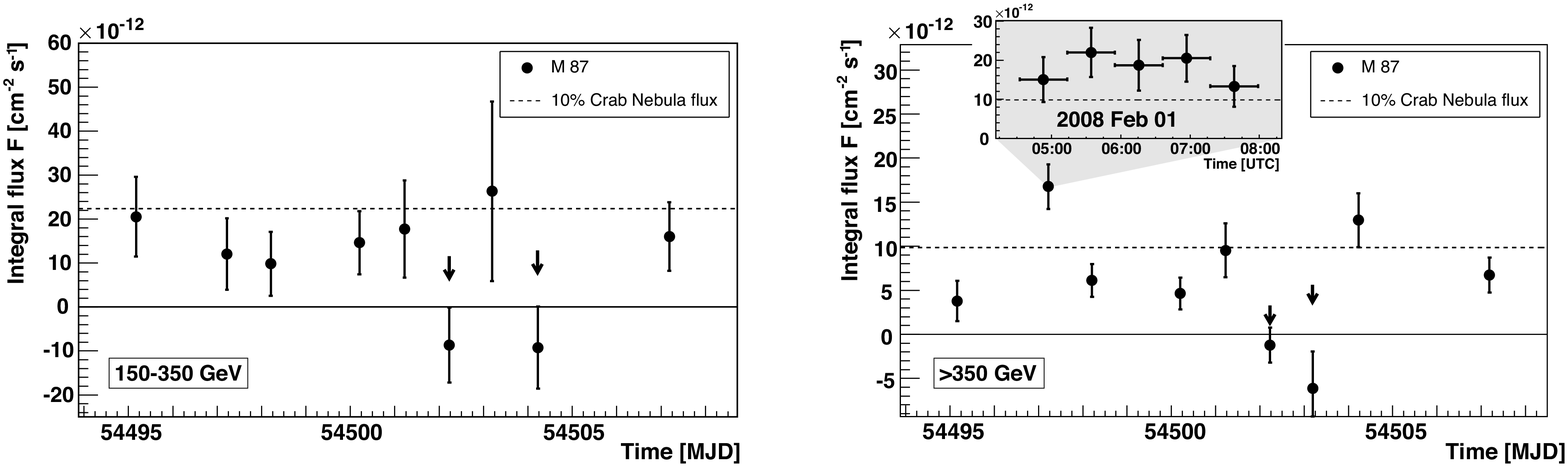}}
\caption{\label{fig:m87} The night-by-night light curve for M\,87
as measured from 2008 January 30 to 2008 February 11. Left: The flux 
in the energy bin $150 - 350$\,GeV is consistent with a constant emission. 
Right: Flux variations are apparent on variability
timescales down to 1 day in the integral flux above 350\,GeV, instead. 
The inset shows the light curve above
350 GeV in a 40 min time binning for the night with the highest
flux (2008 February 1). The vertical arrows represent flux upper
limits (95\% c.l.) for the nights with negative excesses.}
\end{figure}
The giant radio galaxy M\,87 has been known as VHE $\gamma$-ray emitter 
(\citet{hessm87} and refs. therein), and is also one of the best-studied 
extragalactic black-hole systems. To assess variability timescales and the 
location of the VHE engine in M\,87, the H.E.S.S., VERITAS, and MAGIC collaborations
carried out a  shared monitoring of M\,87, resulting in $\approx 120$ h observations
in 2008 \cite{beilicke}. Results from the entire campaign will appear in a 
dedicated paper. During the MAGIC observations, a strong 8\,$\sigma$ signal was
found on 2008 February 1, triggering the other Cherenkov telescopes
as well as {\it Swift} observations. For the first time, MAGIC determined the
energy spectrum below 250 GeV \cite{MAGICM87}, which can
be described by a power law with a spectral index of $\Gamma=2.30
\pm 0.11_\mathrm{stat} \pm 0.20_\mathrm{syst}$.
We did not measure a high-energy cut-off, but found a marginal
spectral hardening, which may be interpreted as a similarity to
other blazars detected at VHE, where such hardening has often been
observed \cite{Wagner3}.
Our analysis revealed a variable (5.6\,$\sigma$)
night-to-night $\gamma$-ray flux above 350 GeV, while no
variability was found for 150--350 GeV
(Fig.~\ref{fig:m87}),  This fastest variability $\Delta t$ observed so far in 
TeV $\gamma$-rays in M\,87, confirming the $E>730\,\mathrm{GeV}$
short-time variability reported earlier \cite{hessm87}, is on the order of or 
even below one day, restricting the emission region to a size of 
$R\leq \Delta t\,c\,\delta = 2.6 \times 10^{15}\,\mathrm{cm} = 2.6\,\delta $ 
Schwarzschild radii (Doppler factor $\delta$), and suggests the
core of M\,87 rather than the brightest known knot in the M\,87 jet,
HST-1, as the origin of the TeV $\gamma$-rays.
During the MAGIC observations
HST-1 was at a historically low X-ray flux level, whereas at the same
time the core luminosity reached a historical maximum (D.
Harris, priv. comm.). This strongly supports the core as
the VHE $\gamma$-ray emission region.

\subsection{The July-2005 Flares of Mkn 501}
MAGIC observed the bright and variable VHE $\gamma$-ray emitter Mkn~501 
during six weeks in summer 2005 \cite{alb07a}. In two of the observation 
nights, the recorded flux ($>4 \times$ that of the Crab nebula) revealed 
rapid changes with doubling times as short as 3 minutes or less. For the 
first time, short ($\approx$~20 min) VHE $\gamma$-ray flares with a 
resolved time structure could be used for detailed studies of particle
acceleration and cooling timescales.

Interestingly the flares in the two nights behave differently:
While the 2005 June 30 flare is only visible in
250\,GeV--1.2\,TeV, the 2005 July 9 flare is apparent in all
energy bands (120~GeV to $>$1.2~TeV). 
Additionally, a photon-by-photon analysis of the July 9 flare \cite{alb07b}
revealed a  time delay between the flare peak at different
energies: At a zero-delay probability of $P=0.026$, a marginal 
time delay of
$\tau_l = (0.030\pm0.012)\,\mathrm{s\,GeV}^{-1}$ towards higher
energies was found using two independent analyses, both exploiting
the full statistical power of the dataset.
Several explanations for this delay have been considered up to
now: (1) Particles inside the emission region moving with constant
Doppler factor need some time to be accelerated to energies that
enable them to produce the highest energy $\gamma$ rays 
\cite{alb07a}. (2) The $\gamma$-ray emission has been captured in 
the initial acceleration phase of the relativistic blob in the jet, 
which at any point in time radiates up to highest $\gamma$-ray energies
possible \cite{bw08}. (3) A one-zone SSC model, which invokes a 
brief episode of increased particle injection at low energies \cite{mas08}.
(4) An energy-dependent speed of photons in vacuum \cite{mat}, as predicted
in some models of quantum gravity \cite{wagnerabano}.
When assuming a
simultaneous emission of the $\gamma$-rays (of different energies)
at the source, a lower limit of $M_{\mathrm{QG}} > 0.21 \times
10^{18}$~GeV (95\% c.l.) can be established \cite{alb07b}.

\subsection{Blazars Detected during Optical Outbursts}
MAGIC has been performing target of opportunity observations upon
high optical states of known or potential VHE blazars. Up to now,
this strategy has proven very successful with the discovery of Mkn
180 \cite{mkn180}, 1ES 1011+496 \cite{MAGIC1011}, and recently 
S5\,0716+71 \cite{atel1500}. The 18.7-h observation of 1ES 1011+496
was triggered by an optical outburst in March 2007, resulting in a 
6.2\,$\sigma$ detection at
$F_{>200\mathrm{GeV}} = (1.58\pm0.32)\times10^{-11}
\mathrm{cm}^{-2} \mathrm{s}^{-1}$ \cite{MAGIC1011}. An indication
for an optical--VHE correlation is given, in that in spring 2007
the VHE $\gamma$-ray flux is >40\% higher than in spring 2006,
where MAGIC observed the blazar as part of a systematic search for
VHE emission from a sample of X-ray bright
($F_{1\,\rm{keV}}$\,$>$\,$2\,\rm{\mu Jy}$) HBLs \cite{systHBL}.
In April 2008, a high optical state of the blazar
S5~0716+71, triggered MAGIC observations, which resulted in a the
detection of a strong 6.8\,$\sigma$ signal in 2.6\,h, corresponding to a flux
of $F_{>400\mathrm{GeV}} \approx 10^{-11} \mathrm{cm}^{-2}
\mathrm{s}^{-1}$ \cite{atel1500}. The source was also in a
high X-ray state \cite{atel1495giommi}.
The determination of the before-unknown redshifts of 1ES\,1011+496
($z=0.21$, \cite{MAGIC1011}) and S5\,0716+71 ($z=0.31$,
\cite{nps08}) makes these objects the third-most and second-most
distant TeV blazars after 3C\,279, respectively.

\subsection{Detection of the flat-spectrum radio quasar 3C\,279}
\begin{figure}
 \resizebox{0.6\columnwidth}{!}
  {\includegraphics{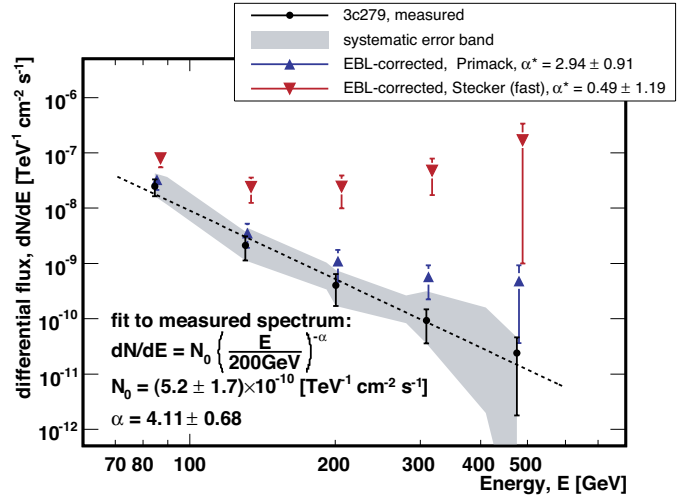}}
\caption{\label{fig:279} Differential energy spectrum of 3C~279.
The grey area includes the combined statistical ($1\sigma$) and
systematic errors. The dotted line shows compatibility of the
measured spectrum with a power law of photon index $\alpha=4.1$.
The triangles are measurements corrected on the basis of two
models for the EBL density (see text). 
}
\end{figure}
Observations of 3C~279, the brightest EGRET AGN \cite{wehrle},
revealed a
5.77\,$\sigma$ post-trial detection on 2006 February 23 supported
by a marginal signal on the preceding night
\cite{MAGICScience}. The overall probability for a
zero-flux lightcurve can be rejected on the 5.04\,$\sigma$ level.
Simultaneous optical $R$-band observations found 3C~279 in a high
optical state, a factor of 2 above its long-term baseline flux,
but with no indication of short time-scale variability. 
The observed VHE spectrum can be described by a power
law with a differential photon spectral index of
$\alpha=4.1\pm0.7_\mathrm{stat}\pm0.2_\mathrm{syst}$ between 75
and 500 GeV (Fig. \ref{fig:279}). The measured integrated flux
above 100~GeV on 23~February is $(5.15\pm 0.82_\mathrm{stat}\pm
1.5_\mathrm{syst}) \times 10^{-10}$ photons cm$^{-2}$ s$^{-1}$.

VHE observations of sources as distant as 3C~279 ($z=0.536$) were
until recently deemed impossible due to the expected strong
attenuation of the $\gamma$-ray flux by the extragalactic background
light (EBL), resulting in an exponential decrease with energy and a
cutoff in the $\gamma$-ray spectrum. The observed VHE spectrum is
sensitive to the EBL between $0.2-2
\mu\mathrm{m}$. The reconstructed intrinsic spectrum of 3C~279 is 
difficult to reconcile with models predicting high EBL densities 
(e.g., the fast-evolution model of \citet{stecker}), while low-level
models, e.g. \cite{primack,ahaebl}, are still viable. Assuming a maximum
intrinsic photon index of $\alpha^\ast = 1.5$, an upper EBL limit
is inferred, leaving a small allowed region for the EBL.

The results support, at higher redshift, the conclusion drawn from
earlier measurements \cite{ahaebl} that the observations of the
{\it Hubble Space Telescope} and {\it Spitzer} correctly estimate
most of the light sources in the Universe.

\section{Galactic Observations} 

In this section we review our latest results on Galactic
astrophysics.

\begin{figure}[!t]
\centering
\includegraphics*[angle=0,height=0.32\columnwidth]{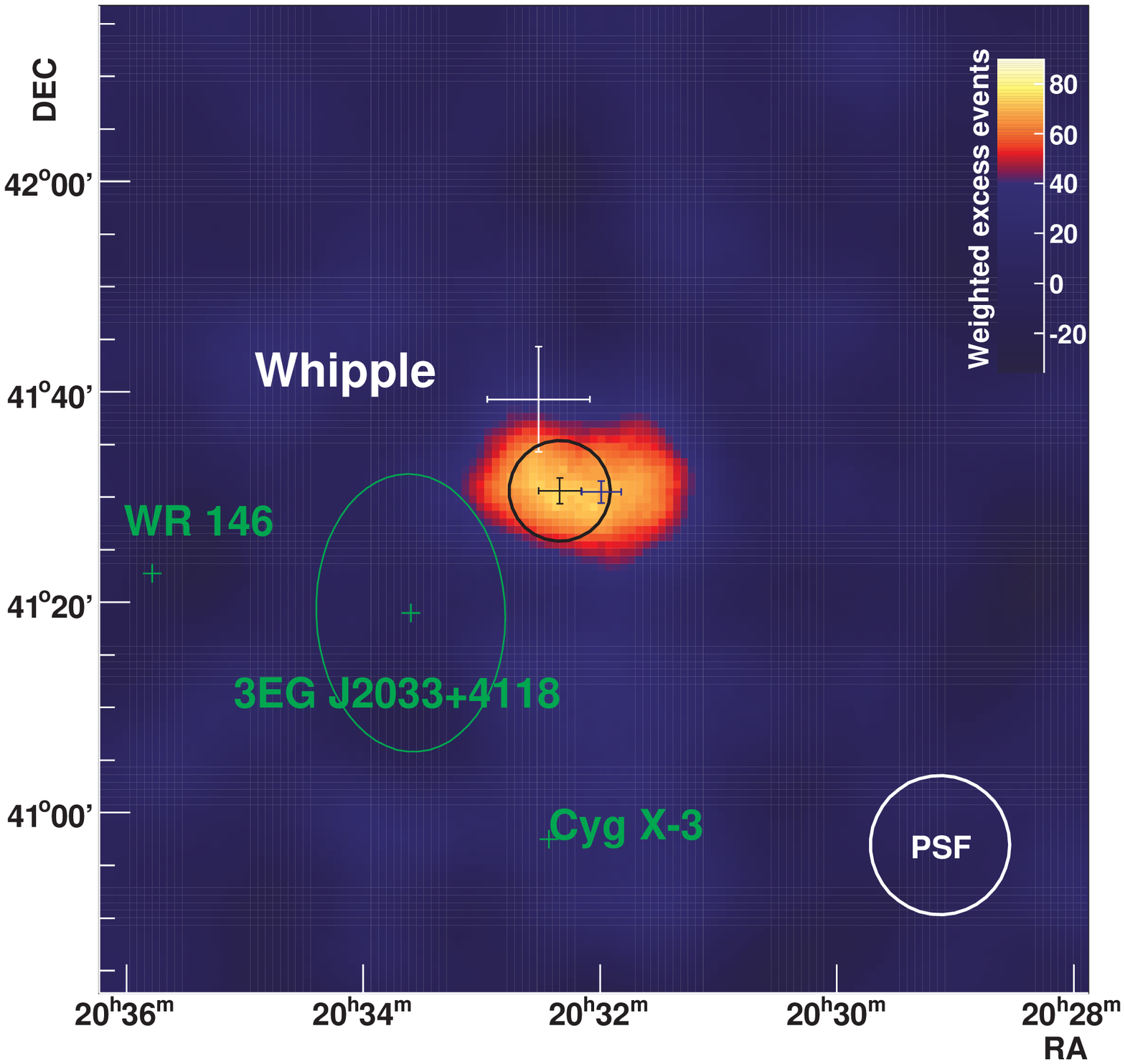}
\includegraphics*[angle=0,height=0.34\columnwidth]{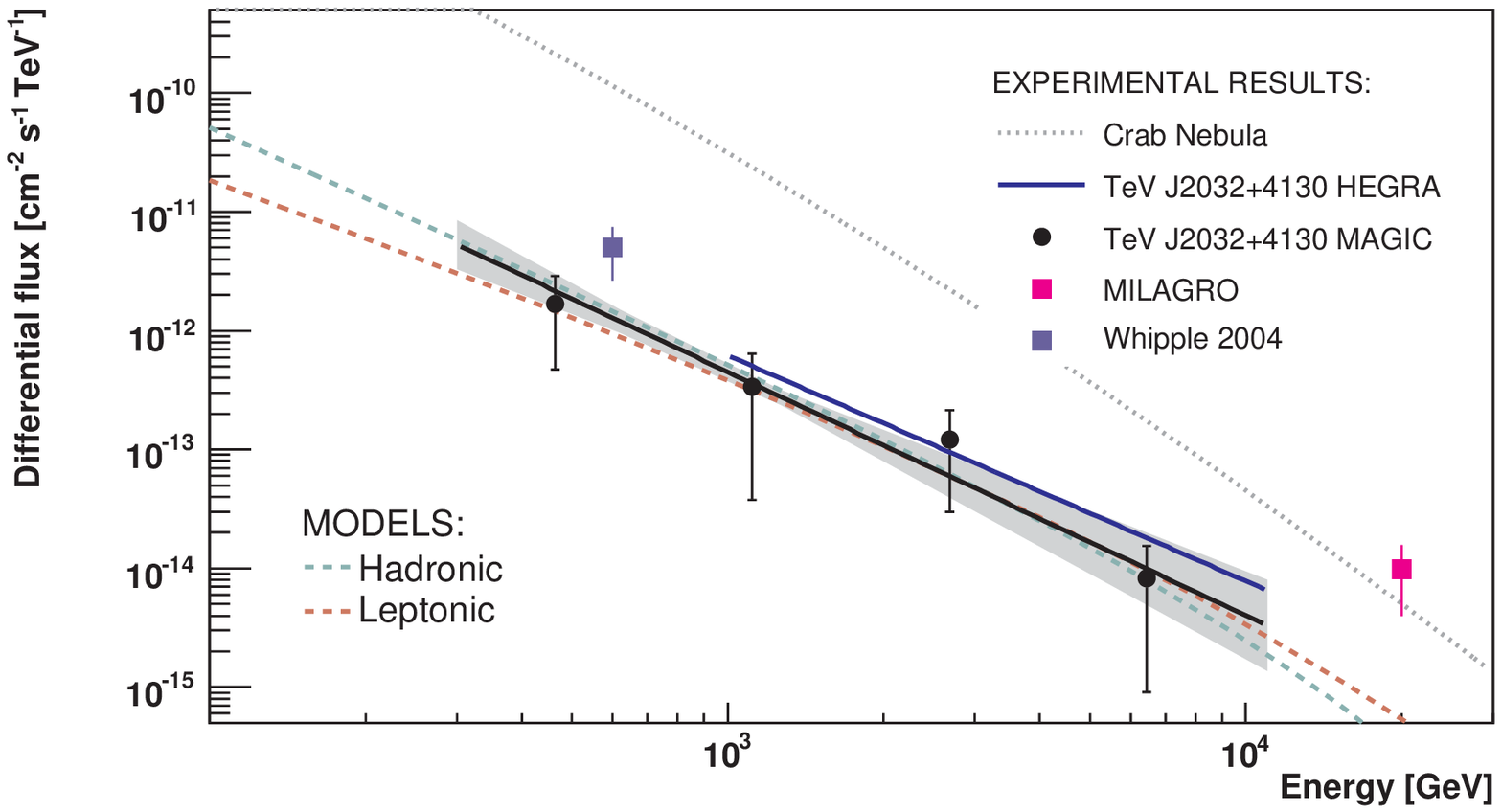}
\caption{Left: Skymap of $\gamma$-ray candidate events
(background-subtracted) around TeV~2032+4132 for energies above
500~GeV. The MAGIC position is shown with a black cross. Also shown
are the last positions reported by Whipple and HEGRA. Right:
Differential energy spectrum from TeV J2032+4130. The shaded area
shows the 1$\sigma$ error in the fitted energy spectrum. The flux
observed by Whipple in 2005 and in the Milagro scan are marked with
squares. The light line shows the HEGRA energy spectrum. Theoretical
one-zone model predictions are depicted with dashed lines}
\label{fig:tev} 
\end{figure}

\subsection{The unidentified $\gamma$-ray source TeV 2032+4130}

The TeV source J2032+4130 was the first unidentified VHE $\gamma$-ray
source, and also the first discovered extended TeV source, likely to
be Galactic~\cite{tevhegra}. The field of view of TeV~J2032+4130 was
observed with MAGIC for 93.7 hours of good-quality data, between 2005
and 2007~\cite{tev}. The source is extended with respect to the MAGIC
PSF (see Figure~\ref{fig:tev}). Its intrinsic size assuming a Gaussian
profile is $\sigma_{\rm src}$ =5.0$\pm$1.7$_{\rm sta}\pm$0.6$_{\rm
sys}$ arcmin. The energy spectrum is well fitted ($\chi^2/n.d.f=0.3$)
by the following power law: $\frac{dN}{dE dA dt} = (4.5\pm
0.3)\times10^{-13}(E/1~TeV)^{-2.0\pm0.3}$
TeV$^{-1}$cm$^{-2}$s$^{-1}$. Quoted errors are statistical, the
systematic error is estimated to be 35$\%$ in the flux level and 0.2
in the photon index~\cite{crab}. The MAGIC energy spectrum (see
Figure~\ref{fig:tev}) is compatible both in flux level and photon
index with the one measured by HEGRA, and extends it down to
400~GeV. We do not find any spectral break, nor any flux variability
over the 3 years of MAGIC observations.

\begin{figure}[t!]
\centering
\includegraphics*[angle=0,height=0.34\columnwidth]{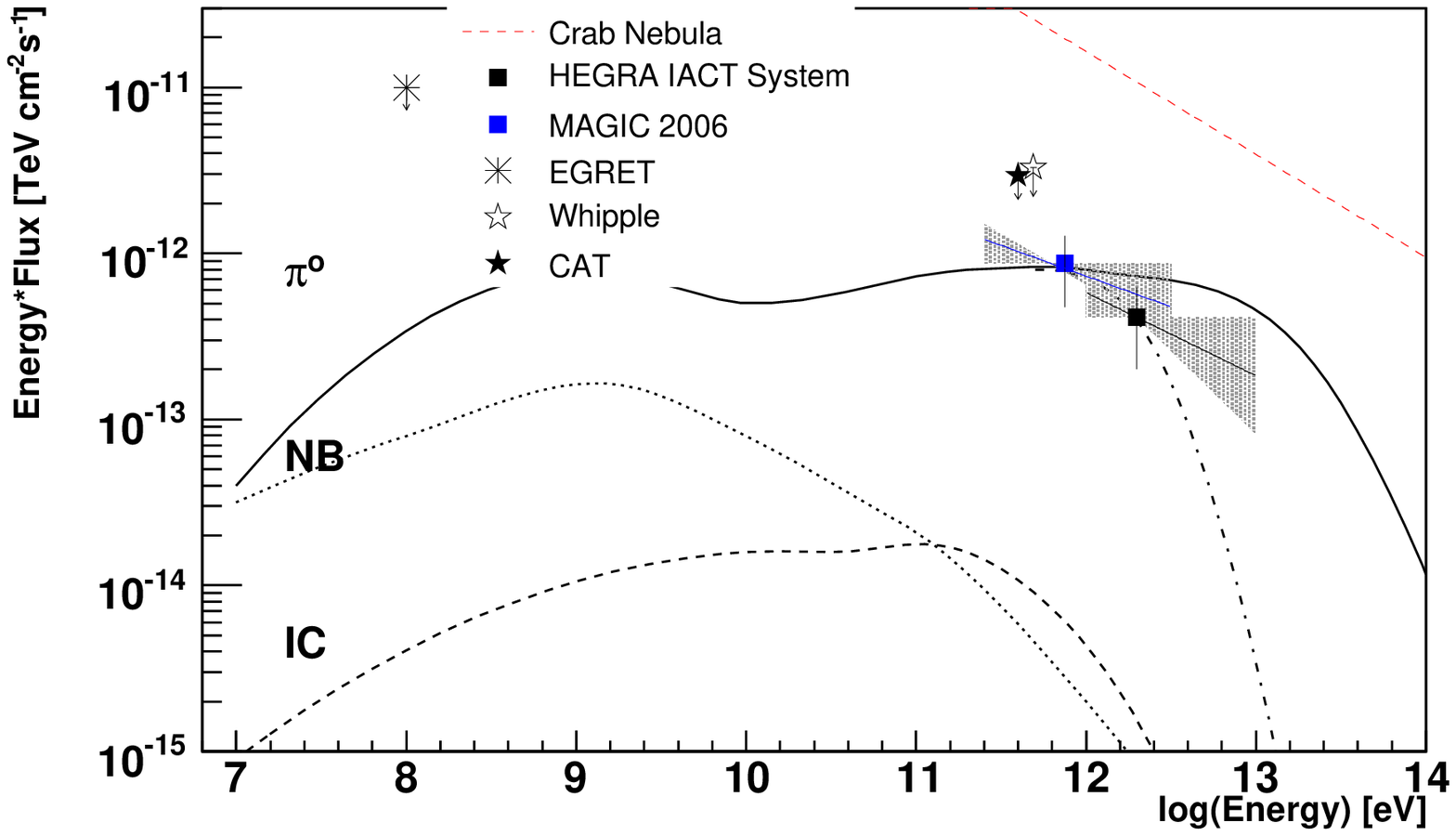}
\includegraphics*[angle=0,height=0.34\columnwidth]{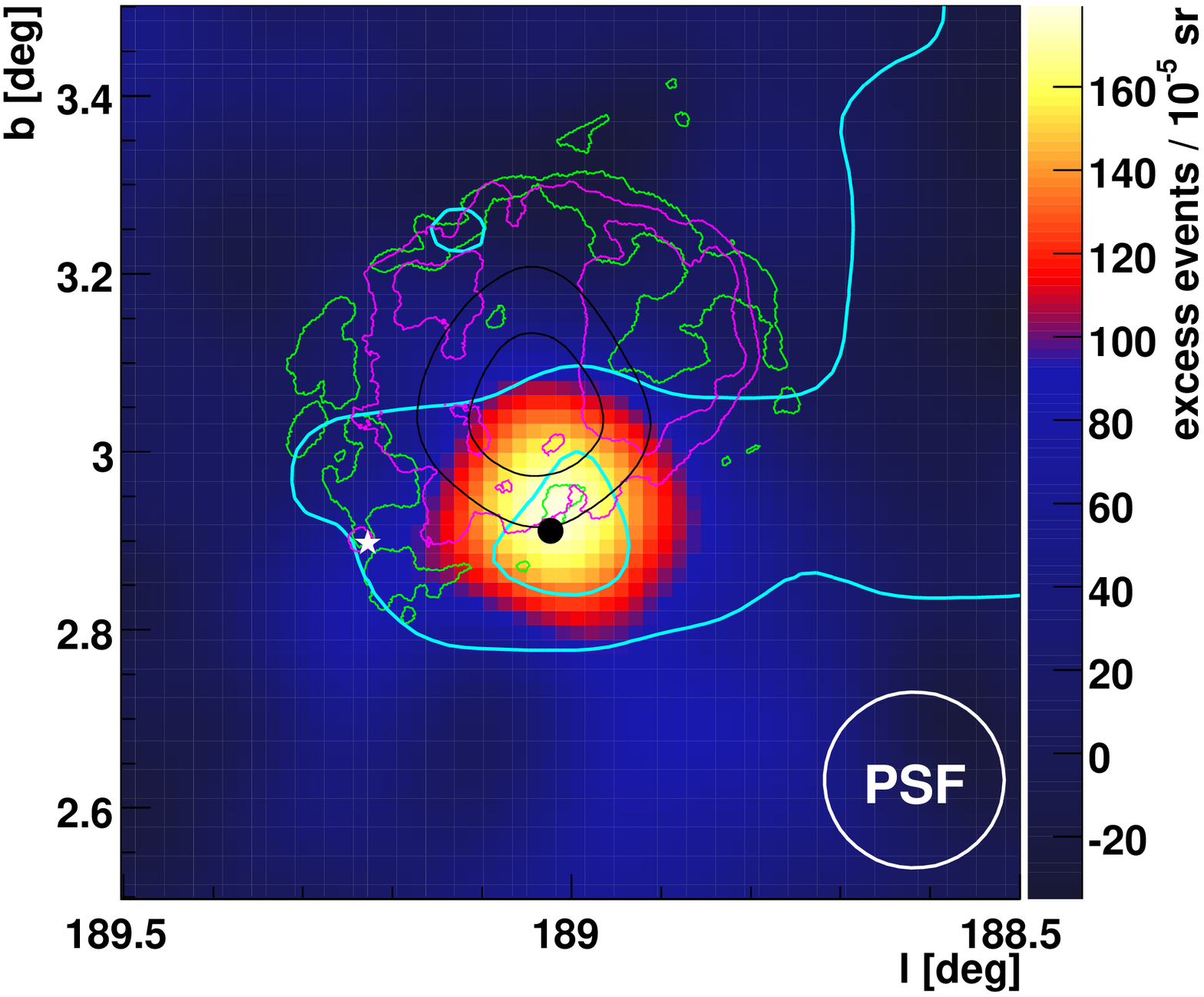}
\caption{Left: Spectrum of Cas~A as measured by MAGIC. The
upper limits given by Whipple, EGRET and CAT are also indicated, as
well as the HEGRA detection. The MAGIC and HEGRA spectra are shown in
the context of the model by \cite{Berezhko}. 
Right: Sky map of $\gamma$-ray candidate events (background
subtracted) in the direction of MAGIC J0616+225 for an energy
threshold of about 150 GeV. Overlayed are $^{12}$CO emission contours,
contours of 20 cm VLA radio data, X-ray contours and $\gamma$-ray
contours from EGRET.  The white star denotes the position of the
pulsar CXOU J061705.3+222127. The black dot shows the position of the
1720 MHz OH maser.}
\label{fig:casA_ic443}
\end{figure}

\subsection{Shell-type Supernova Remnants: Cassiopeia A and IC 433}

We observed the shell-type supernova remnant (SNR) Cassiopeia\,A
during 47 good-quality hours, and detected a point-like source of VHE
$\gamma$-rays above $\sim$250~GeV~\cite{casA}. The measured spectrum
is consistent with a power law with a differential flux at 1~TeV of
(1.0$\pm$0.1$_{stat}\pm$0.3$_{sys})\times10^{-12}$
TeV$^{-1}$cm$^{-1}$s$^{-1}$ and a photon index of
$\Gamma$=2.4$\pm$0.2$_{stat}$$\pm$0.2$_{sys}$. The spectrum measured
about 8 years later by MAGIC is consistent with that measured by
HEGRA~\cite{casAhegra} for the energies above 1~TeV, i.e, where they
overlap (see Figure~\ref{fig:casA_ic443} Left). Our results seem to favor a
hadronic scenario for the $\gamma$-ray production, since a leptonic
origin of the TeV emission would require low magnetic field
intensities, which is in principle difficult to reconcile with the
high values required to explain the rest of the broad-band
spectrum. However, hadronic models~\cite{Berezhko} predict for the 100
GeV -- 1 TeV region a harder spectrum than then measured one.

We have detected a new source of VHE $\gamma$-rays located close to
the Galactic Plane, namely MAGIC\,J0616+225~\cite{ic443}, which is
spatially coincident with the SNR IC\,443. The measured energy
spectrum is well fitted ($\chi^2/n.d.f=1.1$) by the following power
law: $\frac{dN}{dE dA dt} = (1.0\pm
0.2)\times10^{-11}(E/0.4~TeV)^{-3.1\pm0.3}$
TeV$^{-1}$cm$^{-2}$s$^{-1}$. MAGIC\,J0616+225 is point-like for MAGIC
spatial resolution, and appears displaced to the south of the center
of the SNR shell, and correlated with a molecular cloud~\cite{cornett}
and the location of maser emission~\cite{claussen} (see
Figure~\ref{fig:casA_ic443} Right). There is also an EGRET source centered in the
shell of the supernova remnant. The observed VHE radiation may be due
to $\pi^0$-decays from interactions between cosmic rays accelerated in
IC\,443 and the dense molecular cloud. A possible distance of this
cloud from IC\,443 could explain the steepness of the measured VHE
$\gamma$-ray spectrum.

\subsection{Wolf-Rayet binaries}

\begin{figure}[!t]
\centering
\includegraphics*[angle=0,height=0.31\columnwidth]{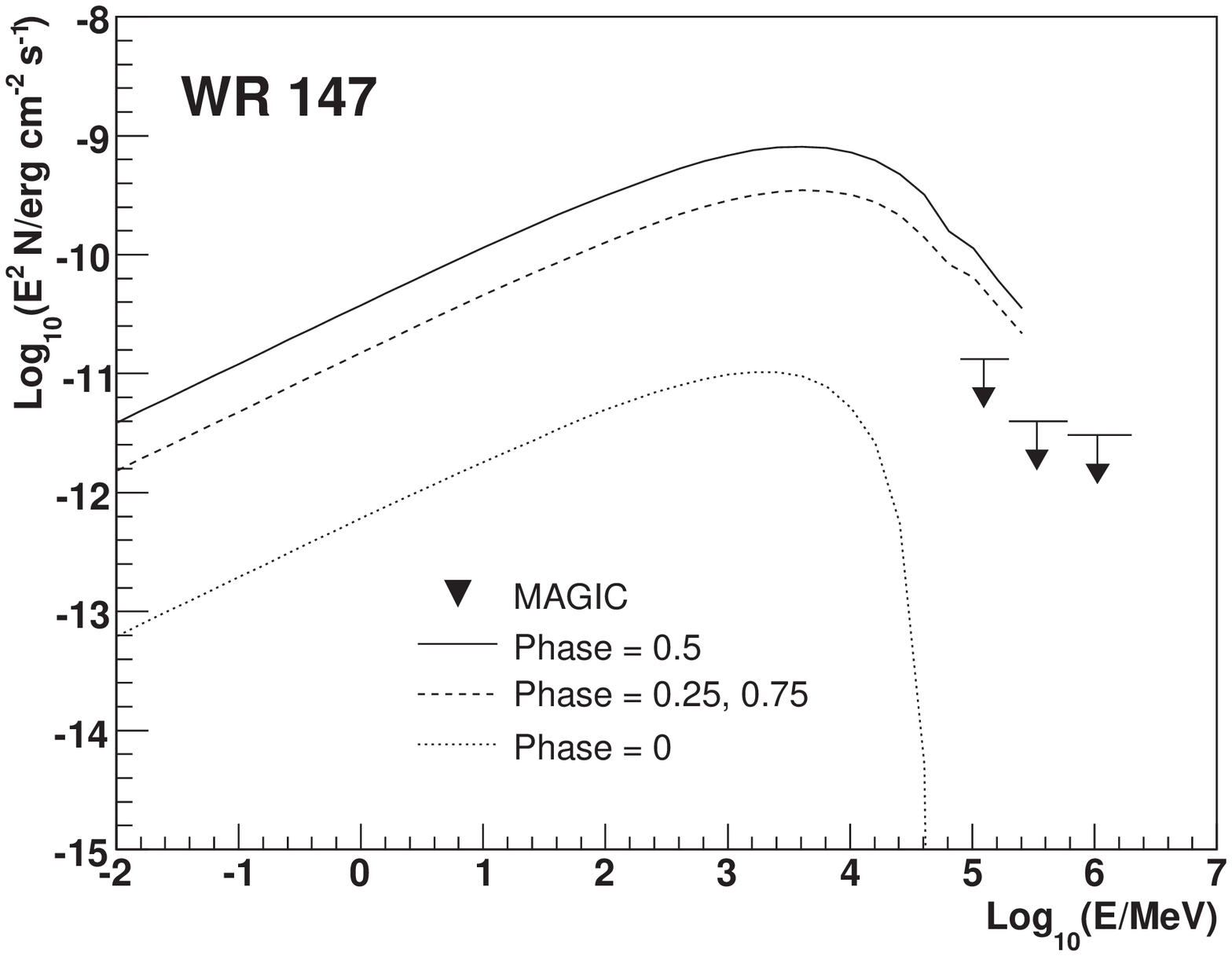}
\hspace{1cm}
\includegraphics*[angle=0,height=0.31\columnwidth]{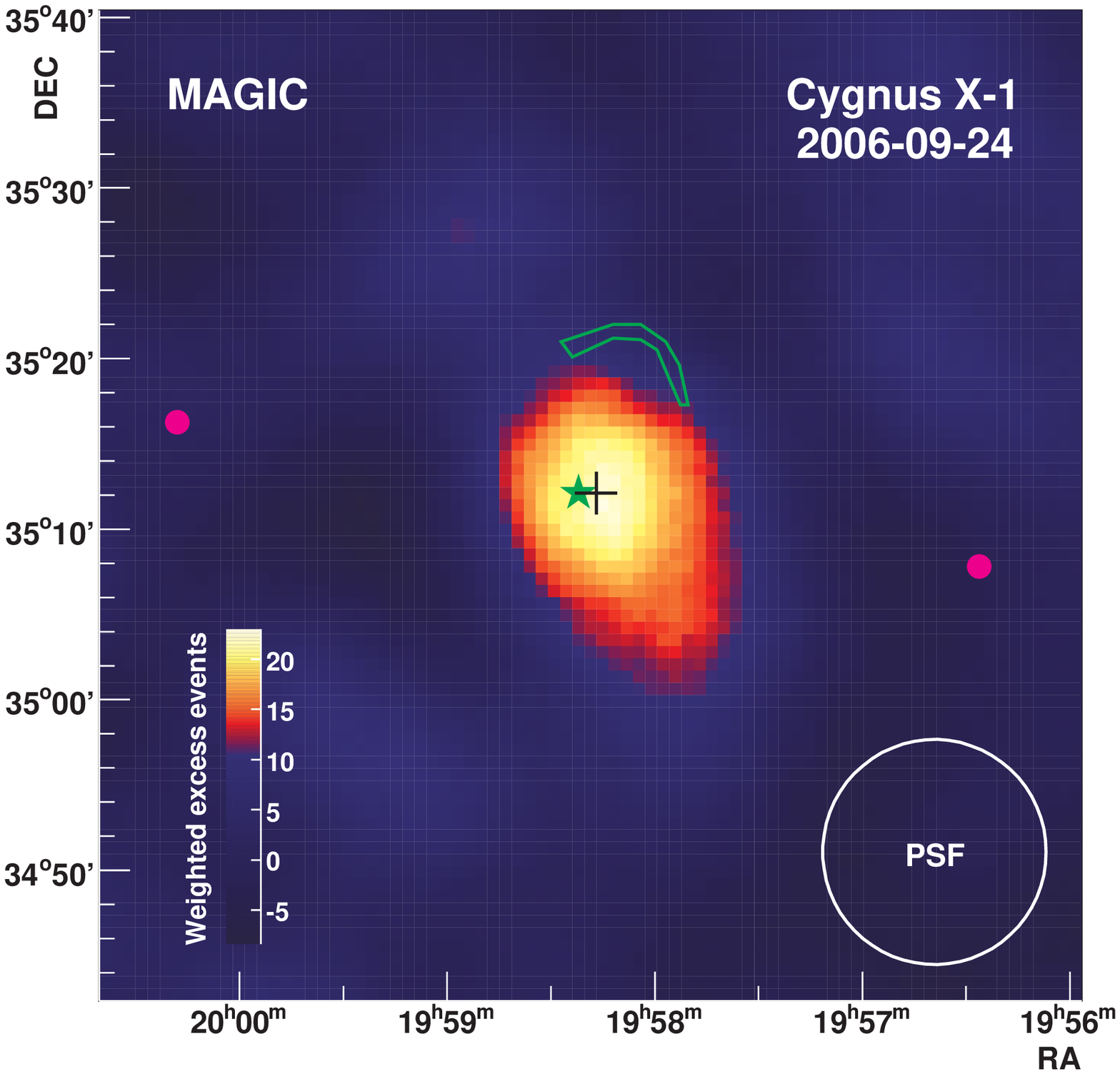}
\caption{\label{fig:wr_cygX1} 
Left: Inverse Compton (IC) spectra of WR\,147 for orbital phases 0,
0.25, 0.5 and 0.75~\cite{Reimer06} together with MAGIC experimental
upper limits.
Right: Skymap of $\gamma$-ray excess events (background subtracted)
above 150 GeV around Cygnus X-1 corresponding to the flare detected on
2006-09-24. The cross shows the best-fit position of the $\gamma$-ray
source. The position of the X-ray source and radio emitting ring-like
are marked by the star and contour, respectively.}
\end{figure}

WR stars display some of the strongest sustained winds among galactic
objects with terminal velocities reaching up to $v_\infty
>1000-5000$\,km/s and also one of the highest known mass loss rate
$\dot M \sim 10^{-4}...10^{-5}~M_\odot/$\,yr.  Colliding winds of
binary systems containing a WR star are considered as potential sites
of non-thermal high-energy photon production, via leptonic and/or
hadronic process after acceleration of primary particles in the
collision shock (see, e.g., \cite{Reimer06}).

We have selected two objects of this kind, namely WR\,147 and WR\,146,
and observed them for 30.3 and 44.5 effective hours,
respectively~\cite{wr147}. No evidence for VHE $\gamma$-ray emission
has been detected in either case, and upper limits to the emission of
1.5, 1.4 and 1.7$\%$ (WR\,147) and 5.0, 3.5 and 1.2$\%$ (WR\,146) of
the Crab Nebula flux are derived for lower energy cuts of 80, 200 and
600 GeV, respectively. These limits are shown in
Figure~\ref{fig:wr_cygX1} (Left) for the case of WR\,147, compared with
a theoretical model~\cite{Reimer06}.

\subsection{Compact binaries: Cygnus X-1 and LS I +61 303}

Cygnus X-1 is the best established candidate for a stellar mass
black-hole (BH) and one of the brightest X-ray sources in the sky. We
have observed it for 40 hours along 26 different nights between June
and November 2006. Our observations have imposed the first limits to
the steady $\gamma$-ray emission from this object, at the level of 1$\%$
of the Crab Nebula flux above $\sim$500\,GeV. We have also obtained a
very strong evidence (4.1\,$\sigma$ post-trial significance) of a
short-lived, intense flaring episode during 24th September 2006, in
coincidence with a historically high flux observed in
X-rays~\cite{malzac} and during the maximum of the $\sim 326$d
super-orbital modulation~\cite{rico}. The detected signal is
point-like, consistent with the position of Cygnus X-1. The nearby 
radio-nebula produced by the jet interaction with the interstellar
medium~\cite{Gallo} is excluded as the possible origin of an eventual
putative emission (see Figure~\ref{fig:wr_cygX1} Right).

\begin{figure}[!t]
\includegraphics[angle=0,height=0.35\columnwidth]{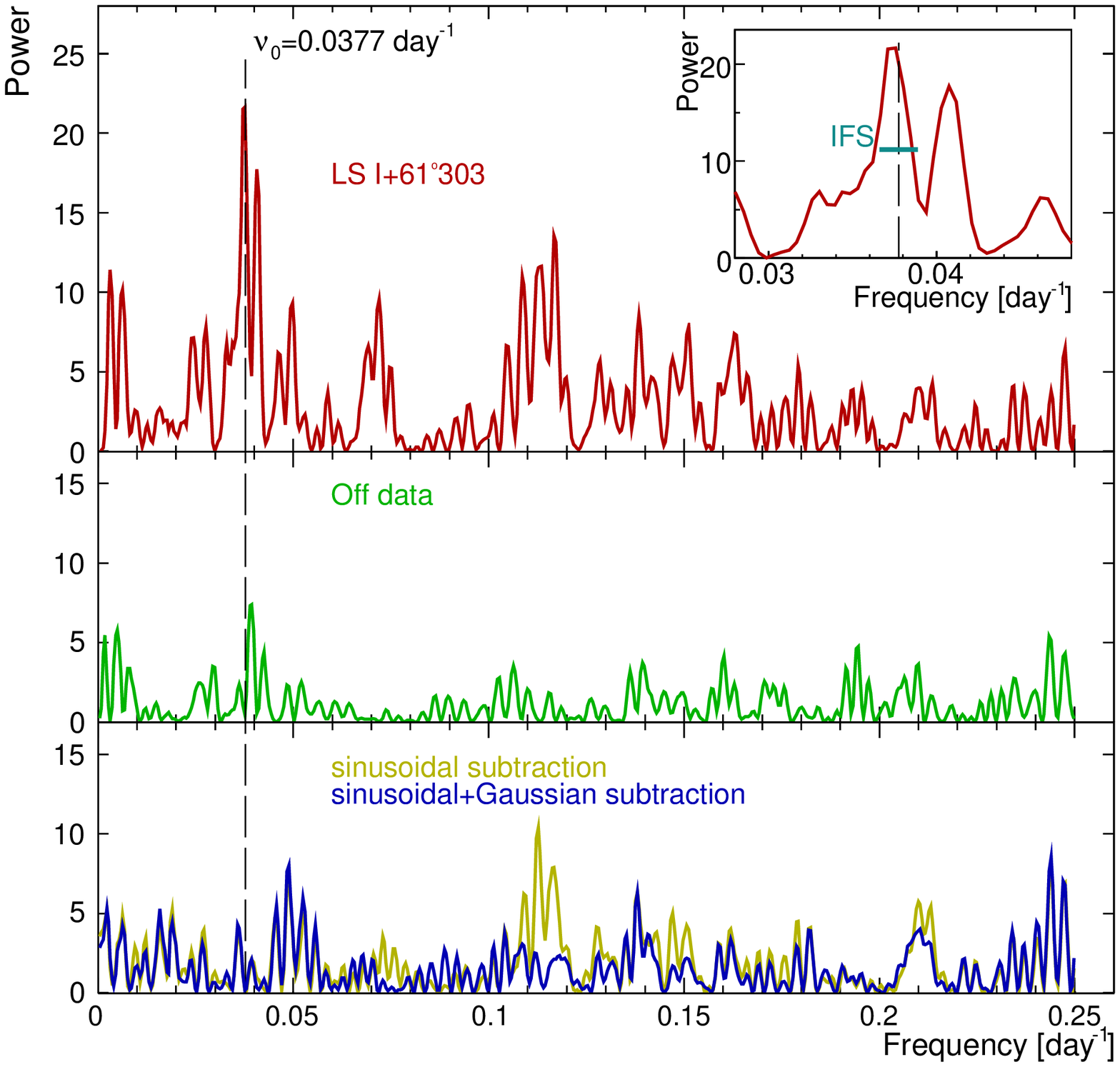}
\hspace{1cm}
\includegraphics[angle=0,height=0.36\columnwidth]{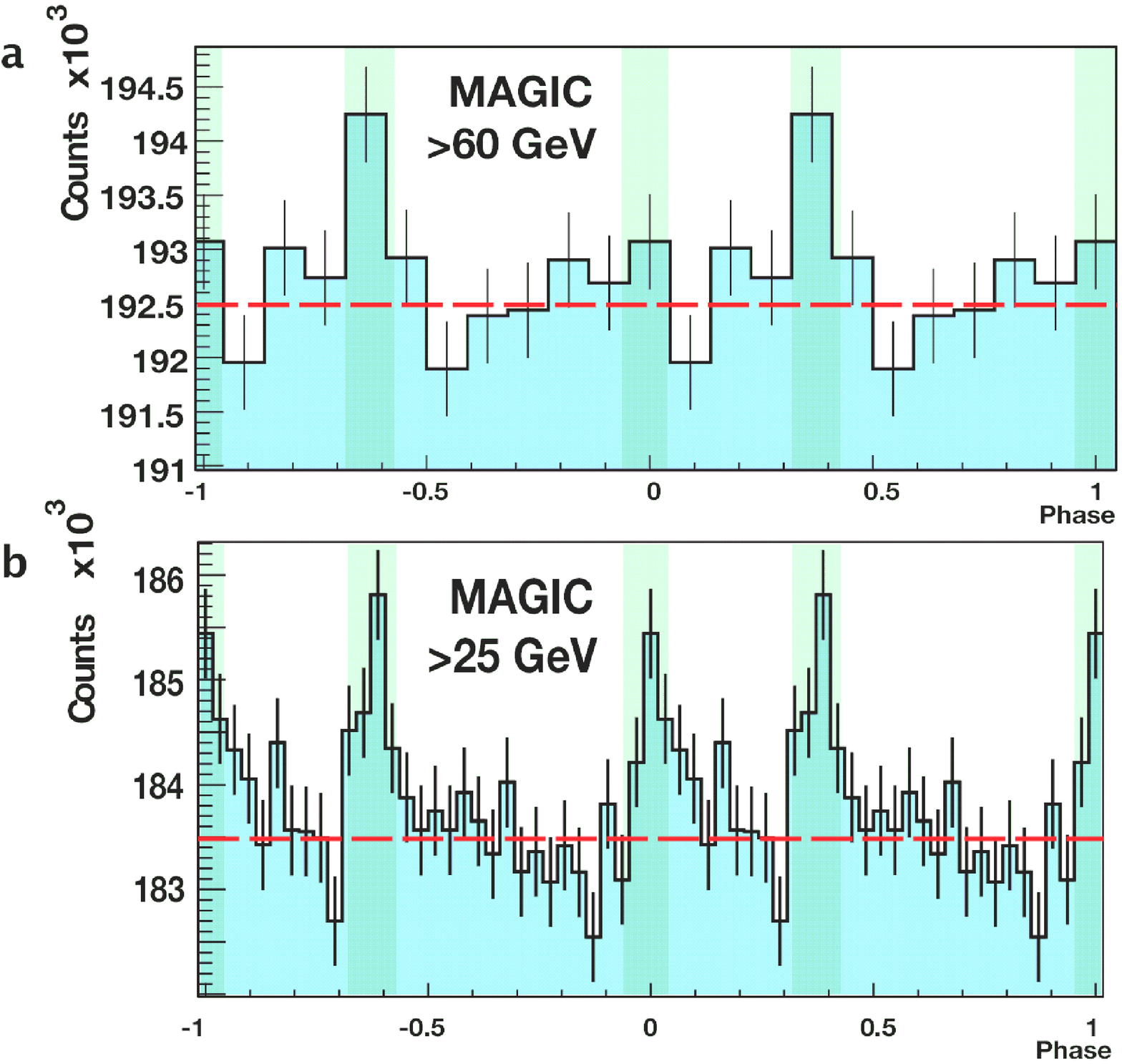}
\caption{Left: Lomb-Scargle periodogram for LS\,I\,+61\,303 data (upper
panel) and simultaneous background data (middle panel). In the lower
panel we show the periodograms after subtraction of a sinusoidal
signal at the orbital period (light line) and a sinusoidal plus a
Gaussian wave form (dark line). The vertical dashed line corresponds
to the orbital frequency. Inset: zoom around the highest peak, which
corresponds to the orbital frequency (0.0377d$^{-1}$). Its post-trial
probability is $\sim$10$^{-7}$. The IFS is also shown. Right: Pulse
profile of the Crab Pulsar above a) 60 GeV and b) 20 GeV, measured by
MAGIC. The shaded areas show the signal regions for the main pulse
(P1) and the inter pulse (P2).
\label{fig:lsi_crab}}
\end{figure}

LS\,I\,+61\,303 is a very peculiar binary system containing a
main-sequence star together with a compact object (neutron star or
black hole), which displays periodic emission throughout the spectrum
from radio to X-ray wavelengths. Observations with MAGIC have
determined that this object produces $\gamma$-rays up to at least $\sim$4
TeV~\cite{lsi}, and that the emission is periodically modulated by the
orbital motion ($P_\textrm{TeV}=(26.8 \pm 0.2)$\,d)~\cite{lsiperiodic}
(see Figure~\ref{fig:lsi_crab} Left). The peak of the emission is found always
at orbital phases around 0.6--0.7. During December 2006 we detected a
secondary peak at phase 0.8--0.9. Between October-November 2006, we
set up a multiwavelenght campaign involving radio (VLBA, e-EVN,
MERLIN), X-ray (Chandra) and TeV (MAGIC) observations~\cite{lsimw}.
We have excluded the existence of large scale ($\sim 100$
mas) persistent radio-jets, found a possible hint of
temporal correlation between the X-ray and TeV emissions and evidence
for radio/TeV non-correlation.

\subsection{Crab Nebula and Pulsar}

The Crab Nebula is the standard candle for VHE astrophysics and as
such, a big fraction of MAGIC observation time is devoted to this
object. Out of it, we have used 16 hours of optimal data to measure
the energy spectrum between 60 GeV and 8 TeV~\cite{crab}. The peak of
the SED has been measured at an energy $E=(77\pm 35)$\,GeV. The VHE
source is point-like and the position coincides with that of the
pulsar. More recently, thanks to a special trigger setup, we have
detected pulsed emission coming from the Crab pulsar above 25\,GeV
(see Figure~\ref{fig:lsi_crab} Right), with a statistical significane
of 6.4\,$\sigma$~\cite{crabpulsar}. This result has revealed a
relatively high energy cutoff, indicating that the emission occurs far
out in the magnetosphere, hence excluding the polar-cap scenario as a
plausible explanation for the high-energy origin. This is also the
first time that a pulsed $\gamma$-ray emission is detected from a
ground-based telescope, and opens the possibility of a detailed study
of the pulsar's energy cutoff, which will help elucidate the mechanism
of high energy radiation in these objects. More details can be found
at~\cite{maxim}.

\vspace{0.5cm}

We thank the Instituto de Astrofisica de Canarias for the excellent
working conditions at the Observatorio del Roque de los Muchachos in
La Palma. 

\end{document}